\documentclass[prb,aps,floatfix, twocolumn]{revtex4} 
\usepackage{graphicx,bm}
\usepackage{subfigure} 
\usepackage{amssymb}
\usepackage{amsmath}
\usepackage{verbatim}
\usepackage{my_symbols}
\usepackage{amsfonts}
\usepackage{graphicx}

\begin{document}

\title{Charge pumping in magnetic tunnel junctions: Scattering theory}

\author{Jiang Xiao,$^{1}$ Gerrit E.W. Bauer,$^{1}$ and Arne Brataas$^{2}$}

\affiliation{$^{1}$Kavli Institute of NanoScience, Delft University of Technology, 2628 CJ
Delft, The Netherlands \\
$^{2}$Department of Physics, Norwegian University of Science and Technology,
NO-7491 Trondheim, Norway}

%\author{Jiang Xiao, Gerrit E.W. Bauer}
%\affiliation{Kavli Institute of NanoScience, Delft University of Technology, 2628 CJ
%Delft, The Netherlands}
%\author{Arne Brataas}
%\affiliation{Department of Physics, Norwegian University of Science and Technology,
%NO-7491 Trondheim, Norway}

\begin{abstract}

We study theoretically the charge transport pumped by magnetization dynamics through epitaxial
FIF and FNIF magnetic tunnel junctions (F: Ferromagnet, I: Insulator, N: Normal metal). We
predict a small but measurable DC pumping voltage under ferromagnetic resonance conditions for
collinear magnetization configurations, which may change sign as function of barrier
parameters. A much larger AC pumping voltage is expected when the magnetizations are at right
angles. Quantum size effects are predicted for an FNIF structure as a function of the normal
layer thickness.

\end{abstract} 
\date{\today} 
\maketitle

\pagestyle{plain}
%+++++++++++++++++++++++++++++++++++++++++++++++++++++++++++++++++++++++++++

%#########################################################################
%\section{Introduction}
%\label{sec:intro}

A magnetic tunnel junction (MTJ) consists of a thin insulating tunnel barrier (I) that
separates two ferromagnetic conducting layers (F) with variable magnetization
direction.\cite{ZT} With a thin normal metal layer inserted next to the barrier, the MTJ
is the only magnetoelectronic structure in which quantum size effects on electron
transport have been detected experimentally.\cite{Yuasa:2002} More importantly, MTJs based
on transition metal alloys and epitaxial MgO barriers \cite{Yuasa:2004, Parkin:2004} are
the core elements of the magnetic random-access memory (MRAM) devices \cite{STTMRAM} that
are operated by the current-induced spin-transfer torque. \cite{Slonczewski:1996,
Berger:1996} 

%Detailed insights into the magnetization dynamics of MTJs have been obtained by studying the
%delicate balance between the STT and an applied magnetic field, as measured by the parametric
%down conversion of an AC voltage bias and periodic resistance modulations caused by
%precessional magnetization states \cite{Tulapurkar:2005} or the power spectrum of the thermal
%or current-induced resistance fluctuations. \cite{Deac} On the theoretical front, only the
%static STT has been studied in detail \cite{Slonczewski:2005, Kalitsov:2006, Theodonis:2006,
%Levy:2006}.

It is known that a moving magnetization of a ferromagnet pumps a spin current into an attached
conductor.\cite{Tserkovnyak:2002a} Spin pumping can be observed indirectly as increased
broadening of ferromagnetic resonance (FMR) spectra.\cite{Mizukami:2002} The spin accumulation
created by spin pumping can be converted into a voltage signal by an analyzing ferromagnetic
contact.\cite{Berger:1999} This process can be divided into two steps: (1) the dynamical
magnetization pumps out a spin current with zero net charge current, (2) the static
magnetization (of the analyzing layer) filters the pumped spin current and gives a charge
current. In the presence of spin-flip scattering, the spin-pumping magnet can generate a
voltage even in an FN bilayer.\cite{Wang:2006, Costache:2006} Spin-pumping by a time-dependent
bulk magnetization texture such as a moving domain wall is also transformed into an
electromotive force. \cite{domainwall} Other experiments on spin-pumping induced voltages have
also been reported. \cite{Azevedo:2005, Saitoh:2006}

Here we present a model study of spin-pumping induced voltages (charge pumping) in MTJs.  Since
the ferromagnets are separated by tunnel barrier, we cannot use the semiclassical
approximations appropriate for metallic structures.\cite{Berger:1999, Brataas:2000,
Tserkovnyak:2005, Wang:2006}  Instead, we present a full quantum mechanical treatment of the
currents in the tunnel barrier by scattering theory. The high quality of MgO tunnel junctions
and the prominence of quantum oscillations observed in FNIF structures (even for alumina
barriers) provide the motivation to concentrate on ballistic structures in which the transverse
Bloch vector is conserved during transport. For a typical MTJ under FMR with cone angle $\theta
= 5^\circ$ at frequency $f = 20$ GHz, we find a DC pumping voltage of $\abs{V_{\rm cp}}\simeq
20$ nV for collinear magnetization configurations or AC voltage with amplitude $\tilde{V}_{\rm
cp} \simeq 0.25~\mu$V for perpendicular configurations. The magnetization dynamics-induced
voltages could give simple and direct access to transport parameters of high-quality MTJs, such
as barrier height, magnetization anisotropies and damping parameters in a non-destructive way.
The polarity of the pumping voltage can be changed by engineering the device parameters, etc.
An oscillating signal as a function of the thickness of the N spacer leads to Fermi surface
calipers that are in tunnel junctions not accessible via the exchange coupling.

%#########################################################################
%\section{Structure \& Method}
%\label{sec:sm}

%=========================================
\begin{figure}[b]
	\includegraphics[width=8.cm]{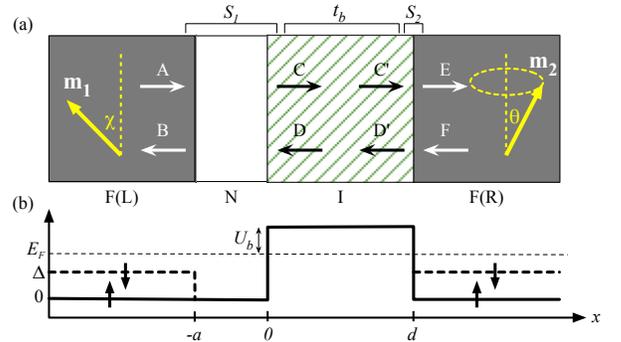}
	 \caption{(Color online) (a): FNIF heterostructure, in which $S_{1,2}$ indicate two
	 different scattering regions. (b): Potential profiles for majority and minority spins in F
	 are shown by solid and dashed lines. The exchange splitting is $\Delta$ and the tunnel
	 barrier has height $U_{\rm b}$ relative to the Fermi energy $\EF$. }
	\label{fig1}
\end{figure}
%=========================================
We consider a structure shown in \Figure{fig1}(a), where two semi-infinite F leads (F(L) and
F(R)) are connected by an insulating layer (I) of width $d$ and a non-magnetic metal layer (N)
of width $a$. The magnetization direction of F(L)/F(R), $\ml/\mr$ ($\abs{\ml} = \abs{\mr} =
1$), is treated as fixed/free. We disregard any spin accumulation in F, thus treat them as
ideal reservoirs in thermal equilibrium. This is allowed when the spin pumping current is much
smaller than the spin-flip rate in the ferromagnet, which is usually a good approximations.
The structure reduces to an FIF MTJ when $a=0$. Let $A, B, \dots, F$ be the spin-dependent
amplitudes ($A^\dagger = (A_\up^\dagger, A_\dn^\dagger)$) at specific points (see
\Figure{fig1}) of {\it flux-normalized} spinor wave-functions.
%:
%%-----------------------------
%\begin{subequations}
%\begin{align}
%	\bra{\psi_\pm^{\rm N}}-i\partial_x\ket{\psi_\pm^{\rm N}} 
%			&= \pm 1 \quad\mbox{in N}, \\
%	\bra{\psi_\pm^{\up,\dn}}-i\partial_x\ket{\psi_\pm^{\up,\dn}} 
%			&= \pm 1 \quad\mbox{in F}, \\
%	\bra{\psi_\mp^{\rm I}}\partial_x\ket{\psi_\pm^{\rm I}} 
%			&= \pm 1 \quad\mbox{in I},
%\end{align}
%\end{subequations}
%%-----------------------------
%where $\up$ and $\dn$ are understood as spin-up and spin-down electrons in F. The $\pm$
%(corresponding to the left and right arrows in \Figure{fig1}) denote the left ($-$) and right
%($+$) going solutions in N and F, and two different exponential solutions in I as indicated in
%\Figure{fig1}.
The scattering states can be expressed in terms of the incoming waves $A$ and $F$, such as: 
%-----------------------------
\begin{equation}
	E = \hat{s}_{\rm EA}A + \hat{s}_{\rm EF}F,
%	\begin{pmatrix} E_\up \\ E_\dn \end{pmatrix} = 
%	\hat{s}_{\rm EA}\begin{pmatrix} A_\up \\ A_\dn \end{pmatrix} +
%	\hat{s}_{\rm EF}\begin{pmatrix} F_\up \\ F_\dn \end{pmatrix}
\end{equation}
%-----------------------------
where $\hat{s}_{\rm EA}$ and $\hat{s}_{\rm EF}$ are $2\times 2$ matrices in spin space and can
be calculated by concatenating the scattering matrices of region $S_{1,2}$ and of the bulk
layer I. To first order of the transmission ($t_{\rm b}$) through the bulk I,  
%-----------------------------
\begin{align}
%	\hat{s}_{\rm EA} = \hat{t}_2 
%	(t_{\rm b} + t_{\rm b}\hat{r}_1'r_{\rm b} + r_{\rm b}'\hat{r}_2t_{\rm b}
%	+ t_{\rm b}\hat{r}_1't_{\rm b}'\hat{r}_2t_{\rm b}+\cdots) \hat{t}_1, \nn
	\hat{s}_{\rm EA} = \hat{t}_2 
	\midb{(1-r'_{\rm b}\hat{r}_2)^{-1}t_{\rm b}(1-\hat{r}_1'r_{\rm b})^{-1}} 
	\hat{t}_1,
\label{eqn:sEA}
\end{align}
%-----------------------------
%%-----------------------------
%\begin{equation}
%	\hat{s}_{\rm EA} = \hat{t}_2t_{\rm b}\hat{M}\hat{t}_1, \quad	
%	\hat{s}_{\rm EF} = \hat{r}_2'+\hat{t}_2\hat{M}'t_{\rm b}\hat{r}_1't_{\rm b}\hat{t}_2', 
%\label{eqn:sEA}
%\end{equation}
%%-----------------------------
%where $\hat{M} = (\hat{1} - \hat{r}_1't_{\rm b}\hat{r}_2t_{\rm b})^{-1}$ and $\hat{M}' =
%(\hat{1} - t_{\rm b}\hat{r}_1't_{\rm b}\hat{r}_2)^{-1}$. 
where $\hat{t}_{1,2}$/$\hat{r}_{1,2}$ are the $2\times 2$ transmission/reflection matrices for
$S_{1,2}$ (see \Figure{fig1}), the hatless $t_{\rm b}/r_{\rm b}$ are the spin-independent
transmission/reflection coefficient for the insulating bulk I. The primed and unprimed version
specify the scattering of electrons emitted coming from the left and right, respectively.  The
reflection coefficient $r_{\rm b}$ is due to the impurity scattering inside the bulk I, and its
magnitude mainly depends on the impurity density in I, especially near the interfaces.  All
scattering coefficients are matrices in the space of transport channels at the Fermi energy
that are labeled by the transverse wave vectors in the leads: ${\bf q, q'}$ (the band index is
suppressed). 

%So in general, a scattering matrix has the following form
%%-----------------------------
%\begin{equation}
%	\hat{s} = 
%	\begin{pmatrix} 
%		s_{\up\up}({\bf q, q'}) & s_{\dn\up}({\bf q, q'}) \\
%		s_{\up\dn}({\bf q, q'}) & s_{\dn\dn}({\bf q, q'}) 
%	\end{pmatrix}.
%\end{equation}
%%-----------------------------

The response to a small applied bias voltage can be written as $J_{\rm c} = G_{\rm c} V$ with
conductance $G_{\rm c}$:
%-----------------------------
\begin{equation}
	G_{\rm c} = \sum_{\bf q, q'}~ g_{\rm c}({\bf q, q'}) \qwith
	g_{\rm c} = {e^2\over h}\Trs{\hat{s}_{\rm EA}\hat{s}_{\rm EA}^\dagger }, 
	\label{eqn:Gc}
\end{equation}
%-----------------------------
where $\Trs{\cdots}$ denotes the spin trace and the summation is over all transverse modes in
the leads at the Fermi level.
%the integration is over the Fermi surface projection.  , and the superscript ``flux'' indicates
%a flux-normalized integral as in \Eq{eqn:int-r}. 
%We have also used $\trs{\hat{s}_{\rm EA}^\dagger\hat{s}_{\rm EA}} = \trs{\hat{s}_{\rm
%BF}^\dagger\hat{s}_{\rm BF}}$  in the definition of $g_{\rm c}$ due to charge conservation. 

%When magnetization changes with time, it is known that current flow even in the absense of
%electrical bias. Three different things here should be distinguished: (1) magnetization
%dynamics ecurrents, this has nothing to do with adiabaticity, (2) when $\hbar\abs{\dmr}\ll\EF$,
%we can make the adiabatic approximation, (3) the constant precession makes computation easy.

When the structure is unbiased but the magnetic configuration is time-dependent, a spin current
is pumped through the structure. \cite{Tserkovnyak:2002a} When the dynamics is slow, $\dmi \ll
\EF/\hbar$, it can be treated by the theory of adiabatic quantum pumping. \cite{BB} We consider
a situation in which the magnetization ($\mr$) of one layer precesses with velocity
$\dot{\phi}$ around the $z$-axis with constant cone angle $\theta$, whereas the other
magnetization ($\ml$) is constant (see \Figure{fig1}). We focus on the charge current that
accompanies the spin pumping:
%with $\mr$ precessing around some axis at a constant cone angle $\theta$ and a precession angle
%$\phi$, one finds that a spin current could be pumped out from the precessing magnetization.
%\cite{Tserkovnyak:2002a}  When the precession frequency $\omega\ll\EF/\hbar$, it can be treated
%by the theory of adiabatic quantum pumping, \cite{Buttiker:1994, Brouwer:1998} Here we
%concentrate not on the spin pumping current, but the charge pumping current:
%(evaluated in the right lead here, but it is the same everywhere):
%-----------------------------
\begin{align}
	J_{\rm cp} &= 
%	\int_{\rm FS}^{\rm flux}{d^2{\bf k}\over(2\pi)^2}{d^2{\bf k'}\over(2\pi)^2}~ 
	\sum_{\bf q, q'}
	j_{\rm cp}({\bf q, q'}), 	
	\label{eqn:jcp}\\
	j_{\rm cp} &={e\dot{\phi}\over 2\pi} \Trs{\im{
	{(\partial_\phi{\hat s}_{\rm EA}})\hat{s}_{\rm EA}^\dagger 
	+ {(\partial_\phi{\hat s}_{\rm EF}})\hat{s}_{\rm EF}^\dagger }}. \nonumber
\end{align}
%-----------------------------
When a DC current is blocked (open circuit), a voltage bias $V_{\rm cp}$ builds up
%-----------------------------
\begin{equation}
	V_{\rm cp}=G_{\rm c}^{-1}J_{\rm cp}. 
	\label{eqn:Vcp}
\end{equation}
%-----------------------------

The discussion above is valid for general scattering matrices that {\it e. g.} include
bulk and interface disorders. In order to derive analytical results, we shall make some
approximations. First of all, we assume that spin is conserved during the scattering,
$\hat{t}_i$ for $S_i$ ($i = 1, 2$, similar for $\hat{r}_i$) is collinear with $\mi$:
\cite{Tserkovnyak:2002a} Expanded in Pauli matrices $\hbmsigma = (\hsigma_x, \hsigma_y,
\hsigma_z)$, $\hat{t}_i = t_i^+ + t_i^-\hbmsigma\cdot\mi$, with $t_i^\pm = (t_i^\up \pm
t_i^\dn)/2$.  $t_{i}^\sigma$ ($\sigma = \up, \dn$) is the transmission amplitude for
spin-up/down electrons with spin quantization axes $\mi$ in the scattering region $S_i$.
In the absence of impurities ($r_{\rm b} = r_{\rm b}' = 0$), \Eq{eqn:sEA} becomes
%-----------------------------
\begin{align}
	\hat{s}_{\rm EA} 
	&=  (t_2^+t_{\rm b}t_1^++t_2^-t_{\rm b}t_1^- \ml\cdot\mr)
	\label{eqn:sEA2}\\
	&+ \hbmsigma\cdot
	\smlb{t_2^+t_{\rm b}t_1^-\ml + t_2^-t_{\rm b}t_1^+\mr -it_2^-t_{\rm b}t_1^-\ml\times\mr}.
	\nonumber
\end{align}
%-----------------------------
Since all hatless quantities in this equation are still matrices in ${\bf k}$-space, such as
$t_2^+ = t_2^+({\bf q, q'})$, the order of $t_2, t_{\rm b}, t_1$ as in \Eq{eqn:sEA} should be
maintained. The $\hat{s}_{\rm EF}$ term in \Eq{eqn:jcp} may be disregarded, because only the
part of $\hat{s}_{\rm EF}$ that depends on both $\ml$ and $\mr$ contributes to $j_{\rm cp}$,
and that part is in higher order of $t_{\rm b}$. 

%%-----------------------------
%\begin{align}
%	\hat{s}_{\rm EA} 
%	&=  (t_{1}^{+}t_{2}^{+}+t_{1}^{-}t_{2}^{-} \ml\cdot\mr)t_{\rm b}
%	\label{eqn:sEA2}\\
%	&+ \hbmsigma\cdot
%	\smlb{t_1^-t_2^+\ml + t_1^+t_2^-\mr -it_1^-t_2^-\ml\times\mr} t_{\rm b}.\nonumber
%\end{align}
%%-----------------------------

Another approximation is the free electron approximation tailored for transition metal
based ferromagnets.\cite{Slonczewski:1989} We assume spherical Fermi surfaces for spin-up
and spin-down electrons (in both F(L) and F(R)) with Fermi wave-vectors $\kF^{\up} =
\sqrt{2m\EF/\hbar^{2}}$ and $\kF^{\dn} = \sqrt{2m(\EF-\Delta)/\hbar^{2}}$, with an
effective electron mass $m$ in F. Electrons in N are assumed to be ideally matched with
the majority electrons in F ($\kF=\kF^{\up}, m_{\rm N} = m$). Let $U_{\rm b}$ and $m_{\rm
b} = \beta m$ be the barrier height of and effective mass in the tunnel barrier. The
adopted potential profile is shown in \Figure{fig1}(b). We assume the transverse
wave-vector ${\bf q}$ to be conserved (${\bf q} = {\bf q}'$) by disregarding any impurity
or interface roughness scattering, which means the scattering matrices ($t_{1,2}^\sigma,
t_{1,2}^\pm, t_{\rm b}$) are diagonal in ${\bf k}$-space. With these approximations, the
double summation in \Eqs{eqn:Gc}{eqn:jcp} is replaced by a single integration over
transverse wave-vectors. The scattering amplitudes $t_i^\sigma$ and $r_i^\sigma$ can be
calculated by matching the flux-normalized wave-functions at the interfaces. The
transmission coefficient in the barrier bulk is the exponential decay: $t_{\rm b} =
e^{-\kappa d}$ with $\kappa = \sqrt{2m_{\rm b}U_{\rm b}/\hbar^2 + q^2}$. Then we obtain
our main result from \Eq{eqn:sEA2}:
%-----------------------------
\begin{subequations}
\label{eqn:ls}
\begin{align}
	g_{\rm c} &={e^2\over 2h}e^{-2\kappa d}
	\smlb{ T_1^+ T_2^+ + T_1^- T_2^- \ml\cdot\mr}, 
	\label{eqn:gc2}\\
	j_{\rm cp} &={e\over2\pi}e^{-2\kappa d}~T_{1}^{-} \ml\cdot \nn
	&\midb{\abs{t_2^-}^{2}(\mr\times\dmr)+\im{t_2^{+*}t_2^-}\dmr},
	\label{eqn:jcp2}
\end{align}
\end{subequations}
%-----------------------------
where $T_{i}^{+}=\abs{t_i^\up}^{2}+\abs{t_i^\dn}^{2}$ is the total transmission probability for
scattering region $S_i$, and $T_{i}^{-} = p_i~T_{i}^{+}=\abs{t_i^\up}^{2} -\abs{t_i^\dn}^{2}$
with polarization $p_i = T_i^-/T_i^+$. 
%%-----------------------------
%\footnote{In \Eq{eqn:jcp2}, we dropped two terms proportional to $\ml\cdot\dmr$, because
%one of them is proportional to the imaginary part of the mixing conductance which is
%nearly zero, and the other is zero for the DC case and is much smaller than other terms
%for the AC case for small angle precessions.}
%%-----------------------------
In \Eq{eqn:jcp2}, The term in the square brakets is the transmitted spin pumping current,
and $T_{1} ^{-}\ml$ represents the filtering by the static layer that converts the spin
into a charge current. 
%to contribution to the charge current vanishes. 
%As all modes are summed over, the contribution from the second
%term remains small due to the self-cancellation among different modes when the current is
%evaluated deep inside F. Therefore, we ignore the second term in \Eq{eqn:jcp2} in the
%later calculation.

%The double integral in \Eq{eqn:Gc} and \Eq{eqn:jcp} reduces to
%the single integral:
%%-----------------------------
%\begin{equation}
%	\int_{\rm FS} d^2{\bf k}
%	= \int {k_{\rm F}^\sigma\over k_x^\sigma}2\pi qdq 
%	\ra \int 2\pi q dq
%	\quad(\sigma=\up,\dn).
%	\label{eqn:int-r}
%\end{equation}
%%-----------------------------
%In the second step, $k_x^\sigma$ is absorbed in the scattering matrix due to the
%flux-normalization of the wave-functions. The $k_{\rm F}^\sigma$ is absorbed by the
%linear-response distortion of the Fermi surface due to the applied bias $V$: $\Delta k^\sigma =
%eV m/\hbar^2 k_{\rm F}^\sigma$ ($eV\ra\hbar\omega$ for pumping situation).

%In linear-response limit:
%%-----------------------------
%\begin{align}
%	{h\over 2e^2}g_{\rm c}
%	&= \half\Trs{\hat{s}_{\rm EA}^\dagger\hat{s}_{\rm EA}}
%	 = \abs{s_{\rm EA}^0}^2 + \abs{s_{\rm EA}^1}^2 + \abs{s_{\rm EA}^2}^2 \nn
%	&+ \abs{s_{\rm EA}^3}^2 \sin^2\theta
%	 + 2\re{s_{\rm EA}^{1\dagger}s_{\rm EA}^2}\cos\theta,
%	\label{eqn:gc}
%\end{align}
%%-----------------------------
%where $\cos\theta=\ml\cdot\mr$. Note $s_{\rm EA}^{0,1,2,3}$ are all $\theta$-dependent.

%#########################################################################
%\section{Results}
%\label{sec:res}
For an Fe/MgO/Fe MTJ: $\kF^{\up} = 1.09$ \r{A}$^{-1}$ and $\kF^{\dn} = 0.42$ \r{A}$^{-1}$ for
Fe,\cite{Slonczewski:1989} and $U_{\rm b} \simeq1$ eV and $\beta = m_{\rm b}/m = 0.4$ for MgO.
\cite{Yuasa:2004, beta} This implies $\EF\simeq4.5$ eV, $\Delta\simeq3.8$ eV $\approx0.85\EF$,
and $U_{\rm b} \approx0.25\EF$ ($t_{\rm b}\ll 1$ when $U_{\rm b}>0.1\EF$ and $d>0.5$ nm). For
an FIF structure ($a = 0$), both $S_1$ and $S_2$ contain only a single F(L)/I (for $S_{1}$) or
I/F(R) (for $S_{2}$) interface. From the potential profile in \Figure{fig1}(b),
%-----------------------------
\begin{equation}
	t_1^\sigma = t_2^\sigma
	= {2\sqrt{ik_x^\sigma\kappa/\beta} \over k_x^\sigma + i\kappa/\beta},
	\label{eqn:t1t2}
\end{equation}
%-----------------------------
for ${k_x^\sigma}^2={\kF^\sigma}^{2}-q^2 > 0$ and zero otherwise. 
%\Eq{eqn:t1t2} implies that
%the transmission probability is maximized when $k_x^\sigma/\kappa = 1$, {\it i.e.} the kinetic
%energy of the electron equals to the relative barrier height. At small $q$, $k_x^\dn/\kappa
%\sim 1$ and $k_x^\up/\kappa > 2$, therefore the transmission for spin down electron is larger,
%this means the polarization inversion $\abs{t_i^\up}^{2} < \abs{t_i^\dn}^{2}$ (or $p_i<0$) for
%small $q$.

%for electrons whose $q$ satisfy
%%-----------------------------
%\begin{equation} 
%	q^{2}<q_{c}^{2}(U_{\rm b}) ={\frac{2m}{\hbar^{2}}}
%	{\frac{\EF(\EF-\Delta)-(m_{\rm b}/m)^2U_{\rm b}^{2}}
%	{2(\EF+(m_{\rm b}/m)U_{\rm b})-\Delta}}.  
%	\label{eqn:qcV}
%\end{equation}
%%-----------------------------

If $\mr$ precesses about an axis that is parallel to $\ml$ ($\chi = 0^\circ$ or
$180^\circ$, see \Figure{fig1}) $\ml\cdot\dmr = 0$ and the second term in \Eq{eqn:jcp2}
vanishes. The dot product $\abs{\ml\cdot(\mr\times\dmr)} = 2\pi f\sin^2\theta$ is
time-independent, thus generates a DC signal. Let us consider an FIF MTJ with barrier
width $d = 0.8$ nm, with $\mr$ precessing around the $z$-axis at frequency $f = 20$ GHz
with cone angle $\theta = 5^\circ$. We find a DC charge pumping voltage over the F leads
$V_{\rm cp} \simeq 15$ nV when $\ml$ is parallel to the precession axis ($\chi = 0^\circ$)
and $V_{\rm cp} \simeq -19$ nV when anti-parallel ($\chi = 180^\circ$). \cite{FNF}
%%-----------------------------
%\footnote{For comparison, the DC voltage for a metallic FNF spin valve under same FMR is
%$V_{\rm cp} \simeq 0.2\mu$V in our calculation.}
%%-----------------------------
$\abs{V_{\rm cp}}$ is higher for the anti-parallel configuration simply because its
resistance is higher. When the precession cone angle $\theta = 10^\circ$,
\cite{Costache:2006b} the DC voltage $\abs{V_{\rm cp}}\simeq 60$ nV, similar to a
previously measured pumping voltage in a metallic junction. \cite{Costache:2006} 

%=========================================
\begin{figure}[t]
	\includegraphics[width=8.5cm]{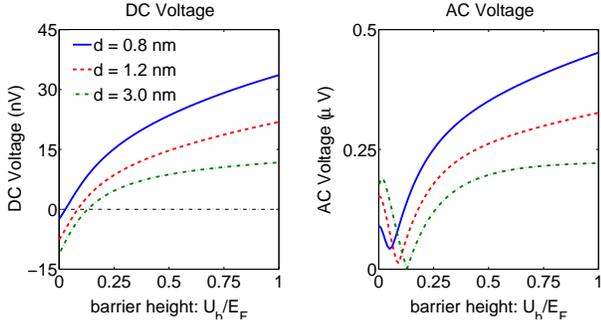}
	\caption{(Color online) Barrier height ($U_{\rm b}$) and width ($d$) dependence of the
	pumping voltage. Left: DC. Right: Maximum amplitude of AC voltage.}
	\label{fig2}
\end{figure} 
%=========================================

\Figure{fig2}(a) shows the DC $V_{\rm cp}$ as a function of the barrier height $U_{\rm b}$
for an FIF structure at $\theta = 5^\circ$ and $f = 20$ GHz. $V_{\rm cp} = G_{\rm c}^{-1}
J_{\rm cp}$ increases as a function of barrier height mainly because $g_{\mathrm{c}} ^{-1}
j_{\mathrm{cp}}$ increases as a function of $U_{\rm b}$: From \Eq{eqn:ls}, we have
$g_{\mathrm{c}} ^{-1} j_{\mathrm{cp}} \approx (T_1^-/T_1^+)(\abs{t_2^-}^2/T_2^+)$ (assume
$T_1^-T_2^- \ll T_1^+T_2^+$). The first ratio $T_1^-/T_1^+ = p_1 \propto
[(\kappa/\beta)^2-k_{\up}k_{\dn}) / [(\kappa/\beta)^2+k_{\up}k_{\dn}]$, increases as a
function of $U_{\rm b}$ through $\kappa(U_{\rm b})$, whereas second ratio
$\abs{t_2^-}^2/T_2^+ \propto (\sqrt{k_\up} -\sqrt{k_\dn})^2/(k_\up+k_\dn)$ is independent
of $\kappa/\beta$. The pumping voltage therefore increases with $U_{\rm b}$ (and
$1/\beta$). We also see that $V_{\rm cp}$ decreases when $d$ increases, which can be
understood by the following: The effect of the tunnel barrier is to focus the transmission
electrons on small $q$'s due to the exponential decay factor $\exp[-2\kappa(q)d]$.
Smaller $q$ implies larger kinetic energy normal to the barrier and therefore reduced
sensitivity to the spin-dependent potentials.  Hence, $V_{\rm cp}$ decreases with barrier
width. The lowest curve in \Figure{fig2}(a) is approximately $V_{\rm cp} = g_{\rm
c}^{-1}(q)j_{\rm cp}(q)|_{q=0}$, because for large $d$ the electrons near $q = 0$
completely dominate the transmission. The negative value of $V_{\rm cp}$ in
\Figure{fig2}(a) is caused by the negative polarization ($p_1 < 0$) at low barrier height
$U_{\rm b}$ for electrons with small $q$. $V_{\rm cp}$ remains finite for infinitely high
or wide barrier, however, the time to build up this voltage, the RC time ($\tau_{RC}$),
goes to infinity due to the exponential growth of the resistance. 

When $\ml$ is perpendicular to the precession axis of $\mr$, {\it i.e.} $\chi = 90^\circ$,
the charge pumping voltage oscillates around zero because both dot products in
\Eq{eqn:jcp2}, $\ml\cdot(\mr\times\dmr) \approx 2\pi f\sin\theta\cos(2\pi f t)$ and
$\ml\cdot\dmr = 2\pi f\sin\theta\sin(2\pi ft)$, give rise to an AC signal. With $V_{\rm
cp} = a~\ml\cdot(\mr\times\dmr) + b~\ml\cdot\dmr$, where the two components are out of
phase by $\pi/2$, the amplitude is given by $\tilde{V}_{\rm cp} = 2\pi
f\sin\theta\sqrt{a^2 + b^2}$.
%%-----------------------------
%\footnote{In this study, we only focus on the term with $\ml\cdot(\mr\times\dmr)$ in
%\Eq{eqn:jcp2}, the second term that is proportional $\ml\cdot\dmr$ is zero in the DC case
%($\ml\cdot\dmr = 0$). For AC case, the second term behaves similarly as the first term but
%with smaller amplitude and $\pi/2$ phase difference. }
%%-----------------------------
An FMR with $\theta = 5^\circ$ and $f = 20$ GHz then gives an AC pumping voltage with
amplitude as large as $\tilde{V}_{\rm cp} \simeq 0.25\mu$V.  \Figure{fig2}(b) shows the
barrier height dependence of amplitude $\tilde{V}_{\rm cp}$ quite similar to the DC case
in \Figure{fig2}(a) and for similar reasons. For a half metallic junction, the magnitude
of the DC pumping voltage $V_{\rm cp}$ can be shown to be bounded by $(\hbar\omega/2e)
\sin^2\theta$ and AC pumping voltage amplitude $\tilde{V}_{\rm cp}$ must be smaller than
$(\hbar\omega/2e) \sin \theta$, where $\omega = 2\pi f$.

\begin{figure}[t]
	\includegraphics[width=8.5cm]{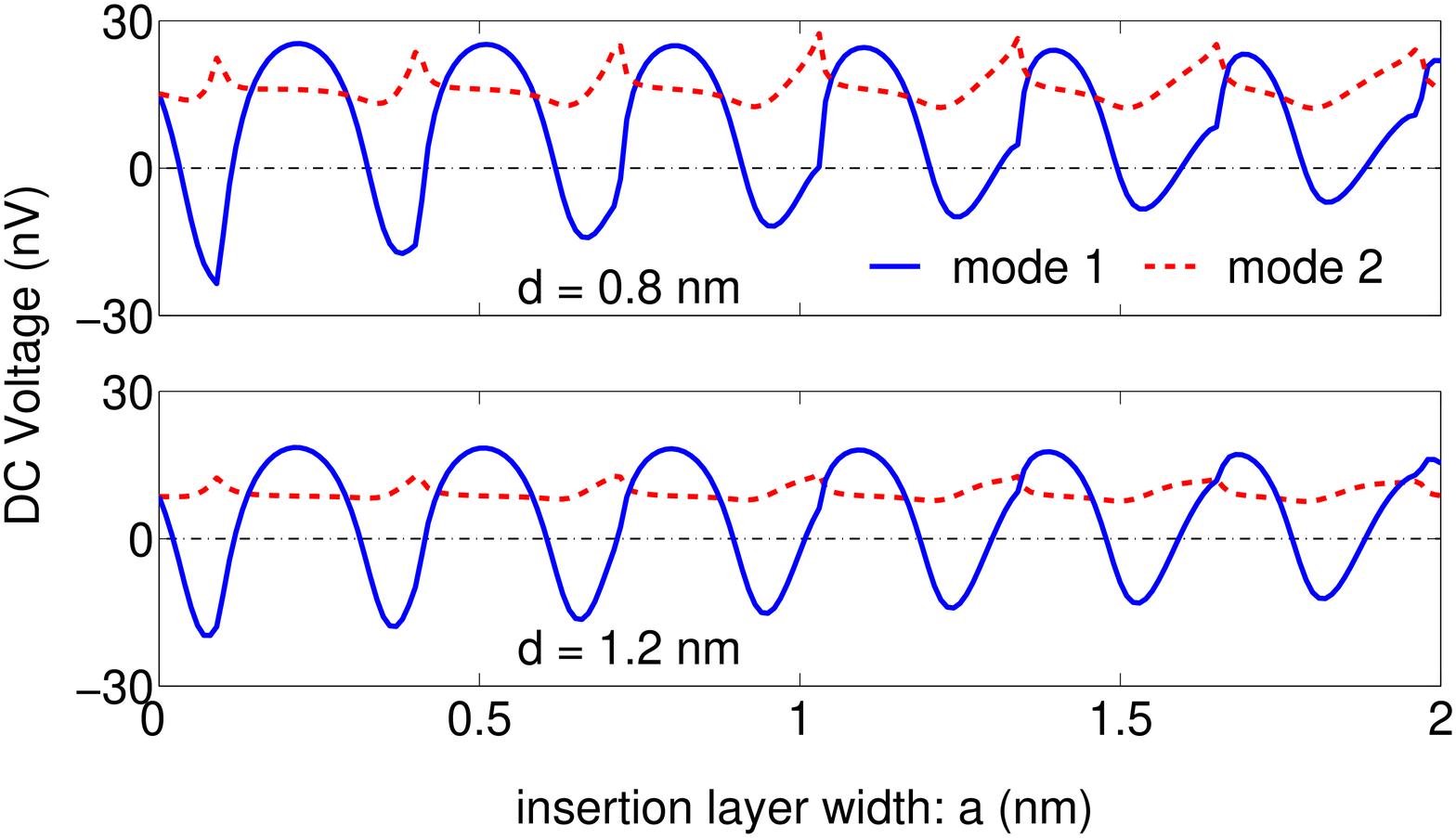}
	\caption{(Color online) $V_{\rm cp}$ vs. N layer thickness $a$ for FNIF ($U_{\rm
	b}=0.25\EF$).}
	\label{Vcp-a}
\end{figure}
%=========================================

When an N layer of thickness $a$ is inserted, it is interesting to inspect the two
different modes, mode 1: FNI$\tilde{\rm F}$ and mode 2: $\tilde{\rm F}$NIF, where
$\tilde{\rm F}$ indicates the F layer under FMR. \Eq{eqn:jcp2} applies to mode 1, and
applies to mode 2 with subscript 1 and 2 swapped. The N layer forms a quantum well for
spin-down electrons that causes the oscillation in the charge pumping voltage as a
function of $a$ as shown in \Figure{Vcp-a}.  The period of the quantum oscillation due to
the N insertion layer is about $\pi/\kF\approx3$ \r{A}. However, due to the aliasing
effect caused by the discrete thickness of the N layer, \cite{Chappert:1991} the observed
period should be $\pi/\abs{\kF - \pi/\lambda}$, where $\lambda$ is the thickness of a
monolayer. In mode 1, the quantum well formed by the N layer can modulate $T_1^-(a)$ such
that the electrons contributing to the transmission the most have $T_1^- > 0$ ($p_1 > 0$)
or $T_1^- < 0$ ($p_1 < 0$), and thus change the sign of the pumping voltage $V_{\rm cp}$.
On the other hand, there is no sign change in mode 2 because $T_2^-$ is independent of
$a$. Similar oscillations could also be found for the amplitude of the AC pumping voltage.

%In \Figure{Vcp-a}, we observe a $V_{\rm cp}$ close to zero at $a = 0$ due to the
%self-cancellation between the large $q$ states with $T_1^- > 0$ and small $q$ states with
%$T_1^- < 0$. 

%For most $a$ values, the magnitude is enhanced by the N insertion layer. 

%#########################################################################
%\section{Discussion} 
%\label{sec:dis}

Because the AC voltage is proportional to $\sin\theta$ and DC voltage is proportional to
$\sin^2\theta$, the AC pumping voltage is much larger than the DC counterpart at small
$\theta$.  However, in order to observe an AC pumping voltage, the time to build up the
voltage, the RC time $\tau_{\rm RC} = RC$, has to be shorter than the pumping period, {\it
i.e.} $\tau_{\rm RC} < 1/f$. Approximately $\tau_{\rm RC}\sim (\varepsilon\varepsilon_0
h/2e^2)e^{2\sqrt{2m_{\rm b}U_{\rm b}/\hbar^2}d}/d$, where $\varepsilon$ and $\varepsilon_0$ are
the dielectric constant and electric constant, respectively. A more accurate estimation of the
RC time for a typical structure is as follows: the resistance-area ($RA$) value of the MTJ in
our calculation is $RA \approx 3~\Omega\mu$m$^2$ for $d = 0.8$ nm ($RA \approx
70~\Omega\mu$m$^2$ for $d = 1.2$ nm, which is consistent with experimental values.
\cite{Yuasa:2004}) The capacitance of an MgO tunnel barrier with $d = 0.8$ nm is calculated by
$C/A = \varepsilon \varepsilon_0/d \approx 0.1$ F/m$^2$ ($\varepsilon \approx 9.7$ for MgO).
Therefore $\tau_{\rm RC} = (RA)(C/A) \approx 0.3$ ps $\ll 1/f \sim 10^2$ ps. The
electromagnetic response is therefore sufficiently fast to follow the AC pumping signal.

We ignored interface roughness and barrier disorder in the calculation of the pumping voltage.
This may be justified by the high quality of epitaxial MgO tunnel barrier.  \cite{Yuasa:2004,
Parkin:2004} Furthermore, the geometric interface roughness mainly reduces the nominal
thickness of the barrier \cite{Zhang:1999} which can be taken care of by an effective thickness
parameter.  Impurity states in the barrier open additional tunneling channels with $U_{\rm
b}^\prime < U_{\rm b}$, which generally increases tunneling but also reduces the spin-dependent
effects when spin-flip is involved. In general, interface roughness and disorder can be
important quantitatively, but have been shown not to qualitatively affect the features
predicted by a ballistic model.  \cite{disorder} In order to be quantitatively reliable, the
real electronic structure has to be taken into account as well.  Both band structure and
disorder effects can be taken into account by first-principles electronic structure calculation
as demonstrated in for metallic structures. \cite{Zwierzycki:2005}

Recently, a magnetization-induced electrical voltage of the order of $\mu$V was measured for an
FIN structure by Moriyama {\it et al}. \cite{Moriyama:2008} The authors explain their findings
by spin pumping, but note that the signal is larger than expected.  An FMR generated electric
voltage generation up to $100\mu$V was theoretically predicted for such FIN structures.
\cite{Chui:2007} Surprisingly, this voltage is much larger than $\hbar\omega/2e \sim \mu$V, the
maximum ``intrinsic ''energy scale in spin-pumping theory. 

To summarize, a scattering matrix theory is used to calculate the charge pumping voltage for a
magnetic multilayer structure. An experimentally accessible charge pumping voltage is found for
an FIF MTJ, the pumping voltage can be either DC or AC depending on the magnetization
configurations. In FNIF structure we find on top of the previously reported oscillating
TMR~\cite{Yuasa:2002} a charge pumping voltage that oscillates and may change sign with the N
layer thickness.

This work has been supported by EC Contract IST-033749 ``DynaMax''.

%#########################################################################
%\bibliography{cpc}

\begin{thebibliography}{35}
\bibitem{FNF}
	\bibinfo{journal}{For comparison, the DC voltage for a metallic FNF spin valve under
	same FMR is $V_{\rm cp} \simeq 0.2\mu$V in our calculation.}
\end{thebibliography}

\end{document}